\documentclass[sn-mathphys-ay]{sn-jnl}
\usepackage{graphicx}%
\usepackage{multirow}%
\usepackage{amsmath,amssymb,amsfonts}%
\usepackage{amsthm}%
\usepackage{mathrsfs}%
\usepackage[title]{appendix}%
\usepackage{xcolor}%
\usepackage{textcomp}%
\usepackage{manyfoot}%
\usepackage{booktabs}%
\usepackage{algorithm}%
\usepackage{algorithmicx}%
\usepackage{algpseudocode}%
\usepackage{listings}%
\usepackage{orcidlink}

\begin{document}

\title{Speed and shape of population fronts with density-dependent diffusion}

\author[1]{\fnm{Beth M.} \sur{Stokes} \orcidlink{0000-0002-7389-053X}} 
\email{bms58@bath.ac.uk}
\equalcont{These authors contributed equally to this work.}

\author[1]{\fnm{Tim} \sur{Rogers} \orcidlink{0000-0002-5733-1658}}
\email{ma3tcr@bath.ac.uk}
\equalcont{These authors contributed equally to this work.}

\author*[2]{\fnm{Richard} \sur{James} \orcidlink{0000-0002-8647-7218}} \email{pysrj@bath.ac.uk}
\equalcont{These authors contributed equally to this work.}

\affil*[1]{\orgdiv{Department of Mathematical Sciences}, \orgname{University of Bath}, \country{UK}}

\affil[2]{\orgdiv{Department of Physics}, \orgname{University of Bath}, \country{UK}}

\abstract {We investigate travelling wave solutions in reaction-diffusion models of animal range expansion in the case that population diffusion is density-dependent. We find that the speed of the selected wave depends critically on the strength of diffusion at low density. For sufficiently large low-density diffusion, the wave propagates at a speed predicted by a simple linear analysis. For small or zero low-density diffusion, the linear analysis is not sufficient, but a variational approach yields exact or approximate expressions for the speed and shape of population fronts.
}


\maketitle

\bmhead{Acknowledgments}
BMS is supported by a scholarship from the EPSRC Centre for Doctoral Training in Statistical Applied Mathematics at Bath (SAMBa), under the project EP/S022945/1. The authors would like to thank Darren Croft and Safi Darden (University of Exeter) for helpful discussions.

\bmhead{Data Accessibility}
This manuscript has no associated data.

\newpage
\section{Introduction}\label{intro}

The Fisher-Kolmogorov-Petrovskii-Piskunov (FKPP) equation \citep{fisher1937wave, kolmogorovstudy}
\begin{equation}
    \frac{\partial u}{\partial t} = D\frac{\partial^2 u}{\partial x^2} + f(u),
    \label{eq.FKPP}
\end{equation}
is the archetypal equation describing the rate of change of a population of density $u(x,t)\in[0,1]$ with time \(t\) in one spatial dimension \(x\) in the presence of diffusion with (constant) coefficient $D$, and density-dependent growth \(f(u)\), with $f(0)=f(1)=0$. The most frequent choice is  logistic growth \(f(u)=ru(1-u)\) with $r$ a parameter fixing the rate of growth. The FKPP equation admits travelling wave solutions $u(x,t)=u(x-ct)$, moving at constant speed $c$  and of fixed profile. It is this feature which makes \eqref{eq.FKPP} a natural starting point for many models of, among other things, the expansion of animal population fronts into new territory  (eg. \cite{murray2002mathematical}).  

If the growth $f(u)$ is logistic, and $D$ a constant, the travelling wave solutions are of the `pulled' type \citep{van2003front}; the wavefront $u(x,t)$ is smooth for small $u$, and the speed of propagation is determined by a linear analysis around the unstable state $u=0$. In brief, setting $u(x,t) = u(z)$ with $z=x-ct$ in \eqref{eq.FKPP} gives $Du''+cu'+f(u)=0,$ which admits travelling wave solutions for all speeds $c \geq c_{L}=2\sqrt{f'(0)D}$. 

Which of the possible wavespeeds is {\it selected} is less straightforward to determine. Numerical solutions of \eqref{eq.FKPP} with $f(u)=ru(1-u)$ and with step-function initial conditions ($u(x,0)=1$ if $x\le x_0$, $0 $ otherwise) will always, modulo numerical error, yield wavefronts moving with speed $c_{L}=2\sqrt{rD}$.  This is as expected; if initial conditions decay faster than a sufficiently steep exponential, the waves have a universal relaxation behaviour to the minimum physically realisable speed as $t \to \infty$ \citep{van2000front}. (This condition is itself a relaxation of the one originally given by \cite{kolmogorovstudy}, that initial conditions with compact support will lead to selection of $c_L$.) Furthermore, any numerical solution of \eqref{eq.FKPP} will necessarily be performed on a finite domain, and all solutions with speed greater than the minimum will be unstable to fluctuations in the moving frame of the wave \citep{canosa1973nonlinear}. So, with some care given to the choice of initial conditions, the solution of the FKPP equation will give rise to travelling waves moving at the {\it minimum realisable speed} $c_{L}=2\sqrt{rD}$. Note that although the speed is known exactly, there is no simple formula for the wave profile $u(z)$.

Most extensions to analysis of  the FKPP \eqref{eq.FKPP} have explored the effect on population dynamics of a range of growth (or reaction) functions $f(u)$. Some of the corresponding PDEs are exactly solvable; an example to which we return has cubic growth $f(u)=u(1-u)(u-a)$, $a\in(0,1/2)$. This case is often referred to as the Nagumo equation, seemingly with reference to a model of neurons \citep{nagumo}. Most of the key features of the linear analysis above carry over to  general $f(u)$ \citep{murray2002mathematical}, providing $f(u)$ has only two zeros, $u_1, u_2$ with $u_1 < u_2$; $f'(u_1) > 0$ and $f'(u_2) < 0$. Usually, as here,  $u_1 = 0$, $u_2 = 1$.

There has been less attention given to the existence and properties of travelling wavefronts when diffusion is density-dependent. Our primary motivation for considering density-dependent $D(u)$ is to model  behavioural aspects of the dispersal of animal populations into new territory. There are many reasons for individuals and populations to undertake movements, such as searching for food or breeding sites, avoiding inbreeding, and in response to man-made factors such as deforestation or climate change. Ecological theorists have predicted that individuals should disperse away from areas of high density and high competition if they can expect to settle in areas of low density and low competition \citep{Poethke, Travis}; in other words, animals move more at higher density to avoid overcrowding. However, empirical support for such positive density-dependent dispersal is mixed \citep{matthysen2005density}. Negative density-dependent dispersal has been observed in various populations, including microorganisms \citep{Jacob}, insects \citep{de-meeus} and fish \citep{de-bona}. Proposed mechanisms for this effect can be broadly categorised as either passive effects due to population structure \citep{matthysen2005density}, or attractive forces which hold back range expansion \citep{bowlerbenton, Travis}.   

Density-dependent diffusion $D(u)$ can be included in the FKPP equation as 
\begin{equation}
    \frac{\partial u}{\partial t} = \frac{\partial}{\partial x} \left[ D(u) \frac{\partial u}{\partial x} \right]+ f(u).
    \label{eq.general_DDD}
\end{equation}
This PDE has been analysed for relatively few choices of diffusion function. It has been analysed in detail for functions of the form  $D(u) = u^m$. The case  $m < 0$ models negative density-dependent diffusion, sometimes referred to as `fast diffusion'. \cite{king2003fisher} investigate the problem for a range of  $m < 0$. They derive approximate analytical solutions, investigate the asymptotic behaviour in both the limits $t \to \infty$ and $m \to 0^-$ and  present numerical simulations. They find that this form of negative density-dependent diffusion leads to \textit{accelerating} wavefronts, not permanent-form travelling wave solutions. 

The case $D(u)=u$ (i.e. $m = 1$), exhibiting positive density-dependence, is exactly solvable  \citep{Aronson_RJ, newman1980some}. With $f=u(1-u$) it yields a travelling wave moving at the unique speed $c = 1/\sqrt{2}$ and with a profile $u(z)=\max\{0,1-\frac{1}{2}\exp(z/\sqrt{2})\}$ - see Fig. \ref{fig:3}.  This case is clearly not amenable to simple linear stability analysis; $u(z)$ is a  \textit{sharp-fronted} travelling wave with discontinuous slope where the density hits $u=0$, and no smooth small-$u$ lead at the front of the wave. \cite{sanchez1994existence, sanchezgarduno1995traveling,sanchez1997travelling} have looked in detail at sharp-fronted travelling wave solutions for different forms of reaction-diffusion equation. In \cite{sanchez1994existence} they show that `degenerate' diffusion functions with $D(0) = 0$ and $D(u) > 0$ for all $u \in (0,1]$ are guaranteed (for appropriate $f(u)$) to produce a sharp-fronted travelling wave.  \cite{Malaguti} show that degenerate ($D(0)=0$) and doubly-degenerate ($D(1)$ also zero) diffusion always leads to a sharp-fronted wave travelling at some unique speed $c^*$ ($=1/\sqrt{2}$ for $D=u$) and to a continuum of smooth waves at faster speeds. \cite{Sherratt}  uses singular perturbation theory to find an asymptotic approximation to  wave solutions travelling faster than $c^*$.

The aim of this paper is to explore the effect of a wider range of diffusion density dependence on travelling wave solutions to \eqref{eq.general_DDD}. Using analytical means where possible, and supported by simple finite-difference numerical simulations, we derive results for the selected speed $c$ of travelling waves, and explore the effect of $D(u)$ on the wave profile $u(z)$. 
In the next section we give an overview of linear and variational methods (following \cite{Hadeler} and  \cite{benguria1996speed}) for computing wave speed in the density-dependent FKPP equation. The remainder of the paper is then arranged around three case studies, employing three different biologically motivated choices of $D(u)$, and illustrating three different classes of behaviour for the selected wavespeed and front shape. 

\section{Methods}\label{methods}
We seek solutions to the FKPP equation with density-dependent diffusion \eqref{eq.general_DDD}. For the most part we will restrict attention to  logistic growth $f=u(1-u)$ (with rate $r=1$), though some of the results we obtain hold for more general growth functions, subject to simple constraints. 

\subsection{Linear analysis}\label{linear}
For some, but not all, diffusion functions $D(u)$ a simple linear stability analysis produces a lower bound $c_L$ on wavespeed, and numerical solutions of \eqref{eq.general_DDD}  select waves travelling at $c_L$, as for the FKPP equation \eqref{eq.FKPP}.  Linearisation of  \eqref{eq.general_DDD} proceeds as follows: set $z=x-ct$ to produce the ODE for $u(z)$
\begin{equation}
    D(u)\frac{\text{d}^2 u}{\text{d}z^2} + \frac{\text{d} D}{\text{d} u} \left( \frac{\text{d} u}{\text{d} z}\right)^2 +c\frac{\text{d} u}{\text{d} z} + f(u)=0.
    \label{eq.ODE_DDD}
\end{equation}
Let  $v = - \text{d} u/\text{d} z$ so that \eqref{eq.ODE_DDD} can be written as the coupled ODE system
\begin{equation}
    \frac{\text{d} u}{\text{d} z} = -v, \quad \frac{\text{d} v}{\text{d} z} = \frac{v^2 D'(u) - cv + f(u)}{D(u)},
    \label{eq.ODE_system}
\end{equation}
where $D'(u)$ indicates a derivative with respect to $u$. The system \eqref{eq.ODE_system} is the starting point for a linear stability analysis of the problem, looking at critical points of trajectories in the $(u,v)$ phase plane; for our choice of variables, trajectories are all in the quadrant where $u$ and $v$ are both non-negative.  

The variable $z=x-ct$ traverses a right-moving wave from left to right, therefore we seek the heteroclinic orbit departing $(u,v)=(1,0)$ and arriving at $(u,v)=(0,0)$. In the neighbourhood of zero, the system \eqref{eq.ODE_system} linearises as
\begin{equation}
    \frac{\text{d}}{\text{d} z}\begin{pmatrix}u\\v\end{pmatrix}=\frac{1}{D(0)}\begin{pmatrix}0&-D(0)\\f'(0)&-c\end{pmatrix}\begin{pmatrix}u\\v\end{pmatrix}\,.
\end{equation}
Depending on the value of $c$, the eigenvalues of this linear system may or may not be real, but complex eigenvalues would imply oscillatory dynamics that are not physically realisable as we do not permit negative population density. It is straightforward to compute the necessary condition that the speed $c$ must exceed $c_L:=2\sqrt{f'(0)D(0)}$. That is, all travelling wave solutions to \eqref{eq.general_DDD} must have speed at least as fast as the \emph{linear speed} $c_L$. For some choices of $D(u)$, however, it is possible that this speed is not physically realisable, and the selected wavefront will in fact travel faster. 

\begin{figure}
    \centering
    \includegraphics[width=0.95\linewidth, trim=100 80 100 80]{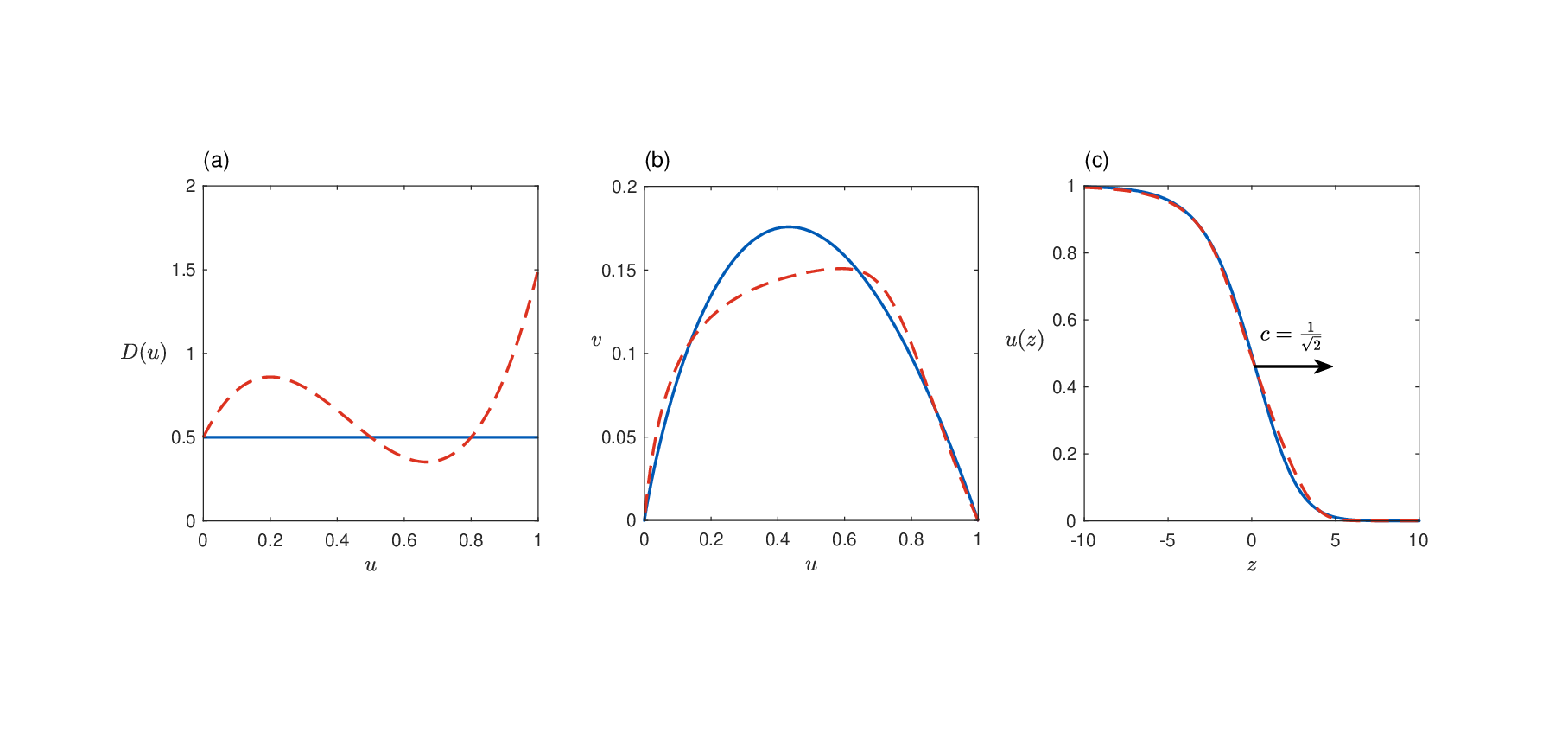}
    \caption{An illustration of solutions to \eqref{eq.general_DDD} for two choices of diffusion function $D(u)$ (a), one constant (solid line) and one with an arbitrarily-shaped density dependence but with $D(0)$ not too small (dashed line). The corresponding $(u,v)$ phase plane trajectories  (b) are clearly quite different, but the  resulting wave profiles (c) are almost identical.  The selected speed for each wave is the linear speed $c_L=1/\sqrt{2}$ in this case.}   \label{fig:1}
\end{figure}

Fig.\ref{fig:1}c shows travelling waves for two diffusion functions that have the same $D(0)=0.5$. They travel at the same speed $c_L=1/\sqrt{2}$ and have almost identical wave profiles  $u(z)$. We find much the same outcome for more-or-less any diffusion function $D(u)$ with $D(0)$ finite and not too small; more examples are found in Section \ref{negative}. For cases where $D(0)$ is small, a different approach is needed.

\subsection{Variational principle}
\label{vp} 


Monotonicity of the wave front implies that we write $v$ as a function of $u$, thereby eliminating explicit dependence on $z$ from the system \eqref{eq.ODE_system} to leave
\begin{equation}
    v \frac{\text{d}}{\text{d} u} \left[ D(u) v\right] - cv + f(u) = 0.
    \label{eq.2a_equiv}
\end{equation}
This ODE for $v(u)$ will be the starting point for all our subsequent analysis of the speed and shape of travelling wave solutions of \eqref{eq.general_DDD}. Unless stated otherwise, $f(u)=u(1-u).$

To derive the variational principle we will closely follow the arguments in 
\cite{benguria1996speed}. For the most part Benguria and Depassier deal with the case of constant diffusion, paying more attention to general growth functions $f(u)$. Though they do not explicitly write down exactly the scheme we use, the key result we need appears in \cite{benguria1996speed} and, in a slightly different form in \cite{Hadeler}.

Let $s(u)$ be any monotonically increasing function with $s(0) = 0$ and $s \to \infty$ as $u \to 1$. 
Multiplying \eqref{eq.2a_equiv} by $\frac{D(u)}{s(u)}$ and integrating by parts with respect to $u$ we obtain
\begin{equation}
    \int_0^1 \frac{f D}{s} du = c \int_0^1 \frac{D v}{s} du - \frac{1}{2} \int_0^1 \frac{s'}{s^2} (Dv)^2 du.
    \label{eq.var_integral}
\end{equation}
Our choice of $s(u)$ means all the integrands in \eqref{eq.var_integral} are positive, so it is useful to define the function 
\begin{equation}
    \phi(v) = c\frac{Dv}{s} - \frac{1}{2} \frac{s'}{s^2} (Dv)^2,
    \label{eq.phi}
\end{equation}
which has a maximum at 
\begin{equation}
    v_{max} = c\frac{s}{s' D}.
    \label{eq.vmax}
\end{equation}
Therefore 
\begin{equation}
    \phi(v) \leq \frac{c^2}{2s'},
    \label{eq.phi_bound}
\end{equation}
for all $v$, and substituting \eqref{eq.phi_bound} into the integrands on the RHS of \eqref{eq.var_integral}, we have that
\begin{equation}
    c^2 \geq 2 \frac{\int_0^1 (fD / s)du}{\int_0^1 (1 / s') du}.
    \label{eq.var_bound}
\end{equation}
This (for a different choice of trial function) is the bound given in \cite{benguria1996speed} for density-dependent diffusion, as an aside to their main result, which considered constant diffusion. Taking the supremum over all trial functions $s(u)$ for which the integrals exist, the maximisation principle is 
\begin{equation}
    \frac{1}{2} c^2 = \sup_{s} \frac{\int_0^1 (fD / s)du}{\int_0^1 (1 / s') du}.
    \label{eq.maxs}
\end{equation}
This is a slightly unusual route to a variational scheme, in that the trial functions $s(u)$ are not trial solutions to the ODE  \eqref{eq.2a_equiv} for $v(u)$. We therefore need also to show that there exists some $\hat{s}(u)$ for which the equality in \eqref{eq.maxs} holds. Then if we can find $\hat{s}(u)$, \eqref{eq.maxs} enables us to find the \textit{exact} wavespeed. From \eqref{eq.vmax}, we know that any such $\hat{s}(u)$ will satisfy
\begin{equation}
    \frac{\text{d} \hat{s}}{\text{d} u} = \frac{c \hat{s}}{Dv},
    \label{eq.s_hat_ODE}
\end{equation}
which can be integrated to obtain (up to a constant)
\begin{equation}
    \hat{s}(u) = \exp\left( \pm c \int \frac{du}{Dv}  \right).
    \label{eq.s_hat}
\end{equation}
If $\hat{s}(u)$ cannot be found, we use the variational principle to find a lower bound on $c$. In passing, we note that for the case where $D$ is constant, expansion of the integrands in \eqref{eq.maxs} about $u=0$ leads to the expected result that $c^2 = 4 f'(0) D(0) = c_L^2.$

Every variational principle has associated with it an Euler-Lagrange equation. For \eqref{eq.maxs} it would be expected that this would be an ODE for $s(u)$, but re-arranging for $u(s)$ yields a more natural form for the associated Euler-Lagrange equation, which becomes
\begin{equation}
    c^2 \frac{\text{d}^2 u}{\text{d} s^2} + \frac{f D}{s^2} = 0.
    \label{eq.euler_lagrange}
\end{equation}
We now have three ordinary differential equations  (\eqref{eq.2a_equiv}, \eqref{eq.s_hat_ODE} and \eqref{eq.euler_lagrange}) all of which offer routes to calculating wavespeed $c$. None of them gives a general route to finding the wave profile; even when $\hat{s}(u)$ is known, there is no guarantee that \eqref{eq.s_hat} can be inverted to find $u(z)$. Note that when $D=1$, \eqref{eq.s_hat} gives $\hat s(u) = \exp(-cz)$, exactly the change of variable needed to transform \eqref{eq.2a_equiv} into \eqref{eq.euler_lagrange} in that case \citep{benguria1996speed}.

One of the potential attractions of this variational approach is that $f(u)$ and $D(u)$ appear in the same term, as a simple product (compare \eqref{eq.euler_lagrange} and \eqref{eq.var_bound} with \eqref{eq.2a_equiv}). Therefore, known results for the FKPP equation \eqref{eq.FKPP} with {\it constant} diffusion and a growth function $f(u)D(u)$ can in principle be exploited to analyse at least the wavespeed of solutions to the density-dependent FKPP equation \eqref{eq.general_DDD} with diffusion $D(u)$ and growth $f(u)$. This fact was recognised and exploited by \cite{Hadeler}, and developed further by \cite{Malaguti2}. We make use of this feature in Section \ref{baseline}.

\section{Case studies}

\subsection{Negative density dependence}\label{negative}

\begin{figure}
    \centering
    \includegraphics[width=\linewidth, trim=100 10 100 10]{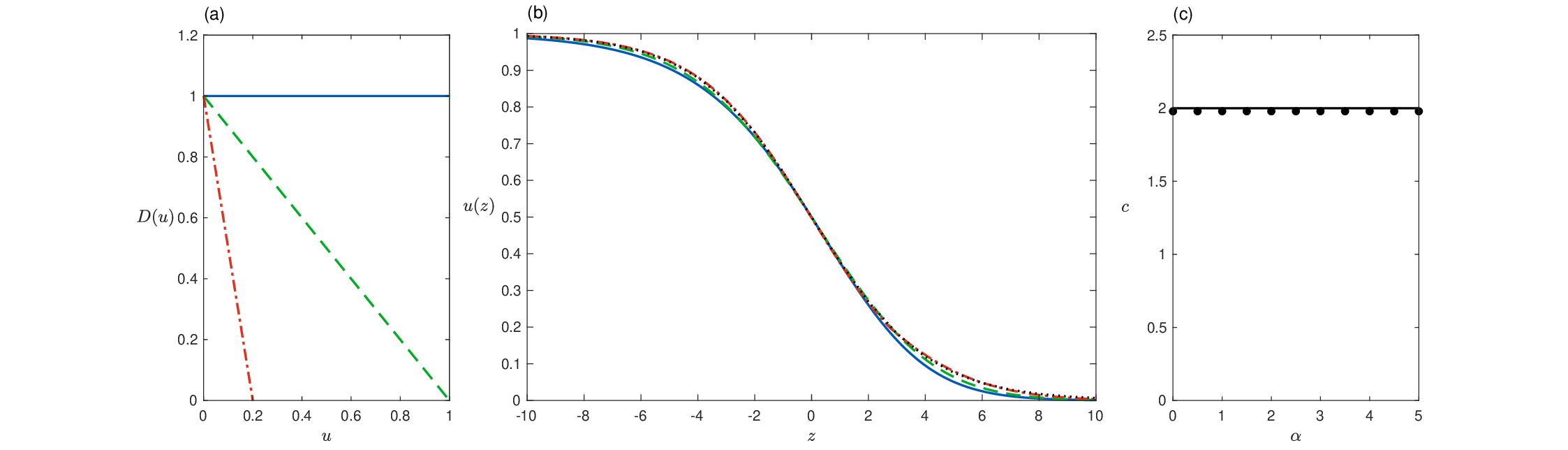}
    \caption{Negative density-dependent diffusion $D=1-\alpha$ (a). The solid line $\alpha=0$ has constant diffusion, for comparison. The other cases in (a) and (b) are $\alpha=1$ (dashed line) and $\alpha=5$ (dot-dashed). The corresponding wave profiles $u(z)$ for these three cases are shown in (b), as is the analytic result for $\alpha\to\infty$ (dotted line). (c) Wavespeed as a function of $\alpha$ - dots are from simulation, solid line is the linear speed $c_L$.}
    \label{fig:2}
\end{figure}

In general, diffusion functions with $D(0)$ sufficiently large will lead to a pulled wave propagating at the linear speed $c_L=2\sqrt{f'(0)D(0)}.$ This is illustrated in Fig.\ref{fig:1}, and again in Fig.\ref{fig:2} for  the simple family of functions $D=1-\alpha u$, $\alpha > 0$. These functions constitute our first case study, exhibiting negative density-dependent diffusion. The growth function is $f=u(1-u)$, so $c_L=2.$  For most choices of $f$ and $D$ there is no analytic expression for the wave profile $u(z)$, so in all case studies we simulate solutions to the PDE \eqref{eq.general_DDD} to generate $u(z)$ and to compare observed speeds to theory. The simulations adopt a simple finite difference scheme. The initial conditions $u(x,0)$ are a step function, and the simulations have a transient period $T=50$ before the profile is recorded and centred on $z=0, u=0.5$. The difference between the predicted and measured wavespeeds (in  Fig.\ref{fig:2}c for example) are within the expected discretisation error of our numerical scheme \citep{van2003front}.

For this set of diffusion functions the shape of the wavefront $u(z)$ hardly varies with $\alpha$, and is always close to the analytic result for $\alpha\to\infty$,  $u(z)=\frac{1}{2}(1-\tanh(z/4))$ (see Fig.\ref{fig:2}b). This result is obtained informally by inserting $c=2$ and $D=0$ in \eqref{eq.2a_equiv} and solving for $v$ and then $u$ via the definition $\frac{du}{dz} = -v$. This limit gives the classic account of a pulled wave - even diffusion  active only at  $u=0$ is enough to populate virgin territory. The state $u=0$ is unstable to population growth, so $f(u)$ then provides an increase in population density up to the (stable) carrying capacity $u=1$. Repetition of these mechanisms pulls along the wave. As noted elsewhere \citep{canosa1973nonlinear} the width of the front, $L$, increases linearly with wavespeed , and therefore scales  here with $\sqrt{D(0)}$. The wave profile (Fig.\ref{fig:2}b) is changed only slightly by the value of  $\alpha$.

Similar results are found for more-or-less any plausible diffusion function with $D(0)$ sufficiently large. Fig.\ref{fig:1} illustrates this for a somewhat arbitrarily chosen diffusion function $D(u)$; the speed of the front is the same as for constant diffusion ($D=D(0)=0.5$ in this figure). We will return to the question what constitutes ``sufficiently large'',  but it is certainly true that all $D(u)$ with a maximum at low density will move at the linear speed $c_L$. To see this, note that one may replace  $D(u)$ by $\max_uD(u)$ in the integrand of \eqref{eq.maxs} to obtain an upper bound on speed which is equal to the linear lower bound in the case $\max_uD(u)=D(0)$. 

Given that in all these cases the travelling waves are clearly pulled, and therefore governed by behaviour at low density, it is perhaps trite to point out that any behavioural mechanism that elevates mean-squared displacement (and hence diffusion) at low density over its values at other densities, would lead in this model to elevated speeds of species invasion or population expansion. We are not aware of this point having been made before; perhaps because the  diffusion functions $D=u^m$ most frequently considered have only the pathological values $D=0$ or $D\to\infty$ at low density, depending on the sign of $m$. For calculations using constant diffusion, there was never a need to be explicit that diffusion at low density is key. Furthermore, in light of the linear stability analysis outlined in Section \ref{linear}, it is perhaps not surprising that diffusion functions with  $D\to\infty$ as $u\to 0$ produce unusual (accelerating) wave solutions \citep{king2003fisher}.

\subsection{Positive density dependence over a baseline}\label{baseline}

\begin{figure}
    \centering
    \includegraphics[width=\linewidth, trim=100 10 100 10]{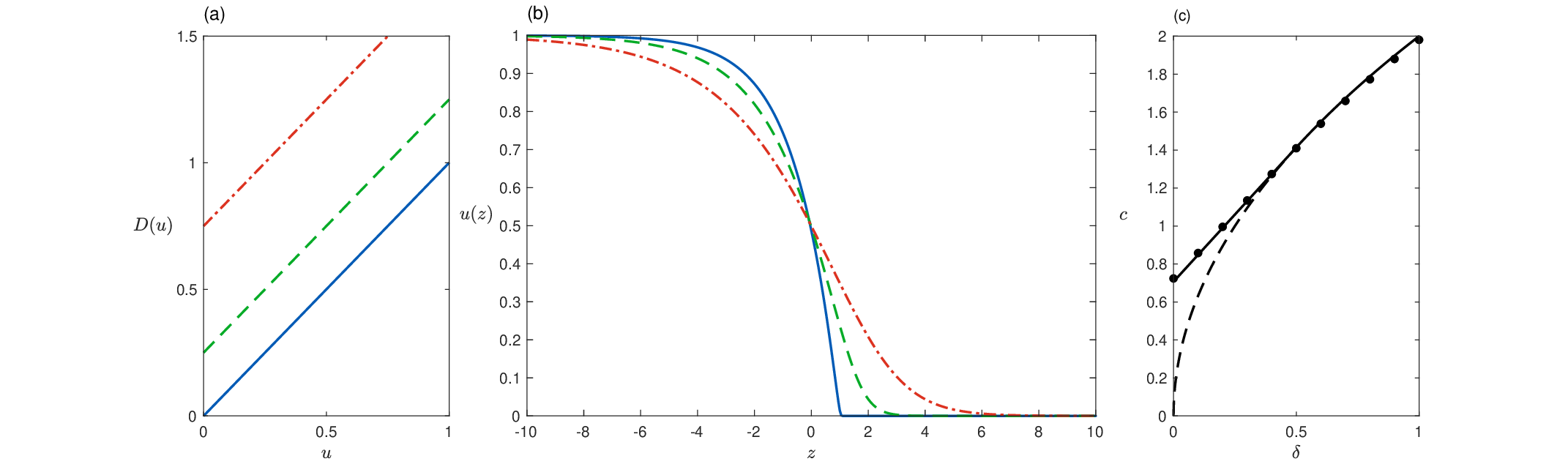}
    \caption{Positive density dependence over a baseline, $D=u+\delta$ (a), with $\delta=0$ (the well-known degenerate diffusion case - solid line), $\delta=0.25$ (dashed line), $\delta=0.75$ (dot-dashed line). (b) The corresponding wavefronts $u(z)$. (c) Wavespeed as a function of $\delta$. Dots are from simulation, solid line is theory \eqref{eq.bound_case1}, dashed line is the linear speed prediction $c_L$.
    }
    \label{fig:3}
\end{figure}

In this section we consider the behaviour of travelling waves with logistic population growth $f(u) = u(1-u)$ and density dependent diffusion  $D(u) = u + \delta$ where $\delta=D(0) \geq 0$ (Fig.\ref{fig:3}a).  Such diffusion functions might correspond to a population response to overcrowding, but with a background level of diffusion at all densities. Then
$$
\frac{\partial u}{\partial t}= \frac{\partial}{\partial x}\left[(u+\delta)\frac{\partial u}{\partial x}\right] + u(1-u)\,.
$$
The presence of the density-dependent term in the diffusion does not prohibit the linear analysis (for $\delta>0$), which gives the linear wavespeed $c_L=2\sqrt{\delta}$. When $\delta=0$ we have the sharp-fronted solution presented in the Introduction, with $c=1/\sqrt{2}$. Here we will reconcile these results; depending on the value of $\delta$, $c_L$ is either the selected speed, or it is not realisable. 

To make use of the variational method (Section \ref{vp}) in practice, it is necessary to identify appropriate trial functions $s(u)$. Ideally there will be a family of functions characterised by a small number of parameters, so that the functional extremisation becomes algebraic. Following \cite{benguria1996speed}, we explore the family
\begin{equation}
    s(u) = \left( \frac{u}{1-u} \right)^{\beta},
    \label{eq.sbeta}
\end{equation}
which has a single (positive) parameter $\beta$. Examining the integrals in \eqref{eq.maxs}, we note that $1/s'$ has a pole of order $1-\beta$ at zero, therefore requiring $\beta<2$ for integrability. Completing the integrals, each of which can be written as a Beta function, yields the following bound:
\begin{equation}
    \frac{1}{2}c^2 \geq \sup_{\beta\in[0,2)} \frac{1}{4}\beta(2-\beta+4\delta)=\begin{cases}
        \displaystyle\frac{(1+2\delta)^2}{4}\quad&\text{if}\quad \delta< 1/2, 
        \\
        2\delta&\text{otherwise.}
    \end{cases}
    \label{eq.bound_case1}
\end{equation}
In the region $\delta\geq\frac{1}{2}$, the bound on the speed matches the linear prediction, and is confirmed by numerical simulations (see Fig.\ref{fig:3}c). More interestingly, when $\delta<\frac{1}{2}$, we have $c\geq (1+2\delta)/\sqrt{2}$, which exceeds the linear prediction. The switch occurs because below $\delta=\frac{1}{2}$ the maximum occurs somewhere in range of allowed $\beta$; beyond this, the maximum is always achieved for the largest possible parameter value, $\beta\to2.$

To show that this bound is tight we must check it is physically realisible. For the present choices of $D$ and $f$, equation \eqref{eq.2a_equiv} becomes
\begin{equation}
    v\big((u+\delta)v'+v-c\big)+u(1-u)=0\,.
    \label{eq.2a_case1}
\end{equation}
Using our candidate $s$ and $c$ in equation \eqref{eq.s_hat_ODE}, we obtain 
$$
v=\frac{cs}{s'(u+\delta)}=\frac{u(1-u)}{\sqrt{2}(u+\delta)}\,,
$$
which by inspection is a non-negative solution to \eqref{eq.2a_case1} with the desired boundary conditions. Note the close parallel with the known solution for the so-called Nagumo equation mentioned in the Introduction. There $D=1$ and the reaction term  $u(1-u)(u - a)$ for $0< a < 1/2$. Apart from the change of sign $(\delta=-a)$, the product of our $D=u+\delta$ and $f=u(1-u)$ is the Nagumo reaction term. The existence of this solution completes the proof that $c=(1+2\delta)/\sqrt{2}$ is the selected speed in this region, since it corresponds to the slowest realisable wave. 

With access to the exact form of $v$, it is also possible to compute the shape of the wavefront when $\delta<\frac{1}{2}$. In the moving frame of reference $z=x-ct$ we have
$$
\frac{\text{d} z}{\text{d} u}=-\frac{1}{v}=-\frac{\sqrt{2}(u+\delta)}{u(1-u)}\,.
$$
Integrating, and choosing $u(0)=\frac{1}{2}$, we obtain  
\begin{equation}
z=\sqrt{2}\Big((1+\delta)\log(1-u)-\delta\log(u)+\log(2)\Big)\,.
\label{eq.zofu}
\end{equation}
Fig.\ref{fig:3}b  shows a selection of these wavefronts. \eqref{eq.zofu} can only be explicitly inverted for special values of $\delta$, including the limiting case $\delta=0$, where we obtain the known sharp-fronted solution $u(z)=\max\{0,1-\frac{1}{2}\exp(z/\sqrt{2})\}$. Finally, note that for the example with $\delta=0.25$ the front has the appearance of a pulled wave, with a tail that might induce a (false) prediction that its wavespeed should be the linear speed $c_L=2\sqrt{\delta}.$ These wavefront solutions are consistent with the smooth wavefronts reported by \cite{Sherratt}, who added a perturbation $\epsilon$ to the wavespeed $c^*=1/\sqrt{2}$. Since speed is proportional to $\delta$ for $\delta <0$, Sherratt's $\epsilon=\sqrt{2}\delta$.

\subsection{Positive density dependence above a threshold}\label{threshold}

\begin{figure}
    \centering
    \includegraphics[width=\linewidth, trim=100 10 100 10]{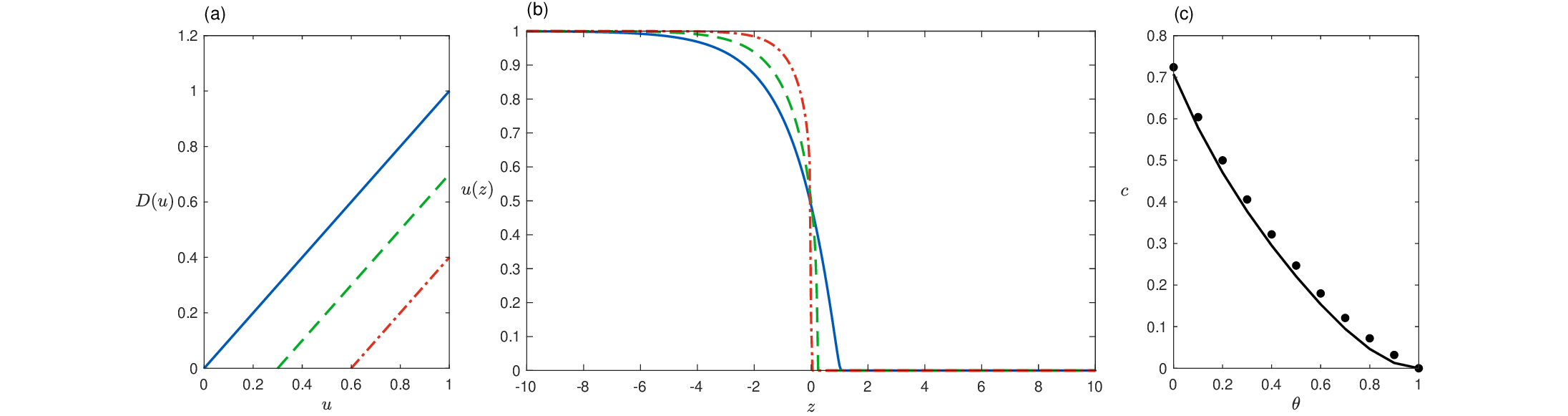}
    \caption{Positive density dependence above a threshold $D(u)=\max\{0, u -\theta\}$ (a), with with $\theta=0$ (the degenerate diffusion case - solid line), $\theta=0.3$ (dashed line), $\theta=0.6$ (dot-dashed line). (b) The corresponding wavefronts $u(z)$ from simulation. (c) Wavespeed as a function of $\theta$. Dots are from simulation, solid line is the approximation \eqref{eq.csimple} from theory \eqref{eq.bound_case1}.    
    }
    \label{fig:4}
\end{figure}

Finally, we will consider the diffusion function $D(u)=\max\{0, u -\theta\}$ (Fig.\ref{fig:4}a). We again have diffusion that increases linearly with density, but only now once the density has exceeded some critical value $\theta$. Biologically we have in mind that populations respond to overcrowding, but only once a threshold has been breached. Mathematically this case is of interest because prohibiting diffusion at small $u$ certainly precludes pulled wavefronts with a small-$u$ tail propagating at linear speed $c_L$. Fig.\ref{fig:4} shows examples of $D(u)$, $u(z)$ for this case, and how wavespeed $c$ varies with threshold $\theta$.

In this case, the family of trial functions used previously \eqref{eq.sbeta} does not yield a simple form from which the supremum in $\beta$ can easily be taken, and nor have we found a different family of trial functions that does the job.

\begin{figure}
    \centering
    \includegraphics[width=\linewidth]{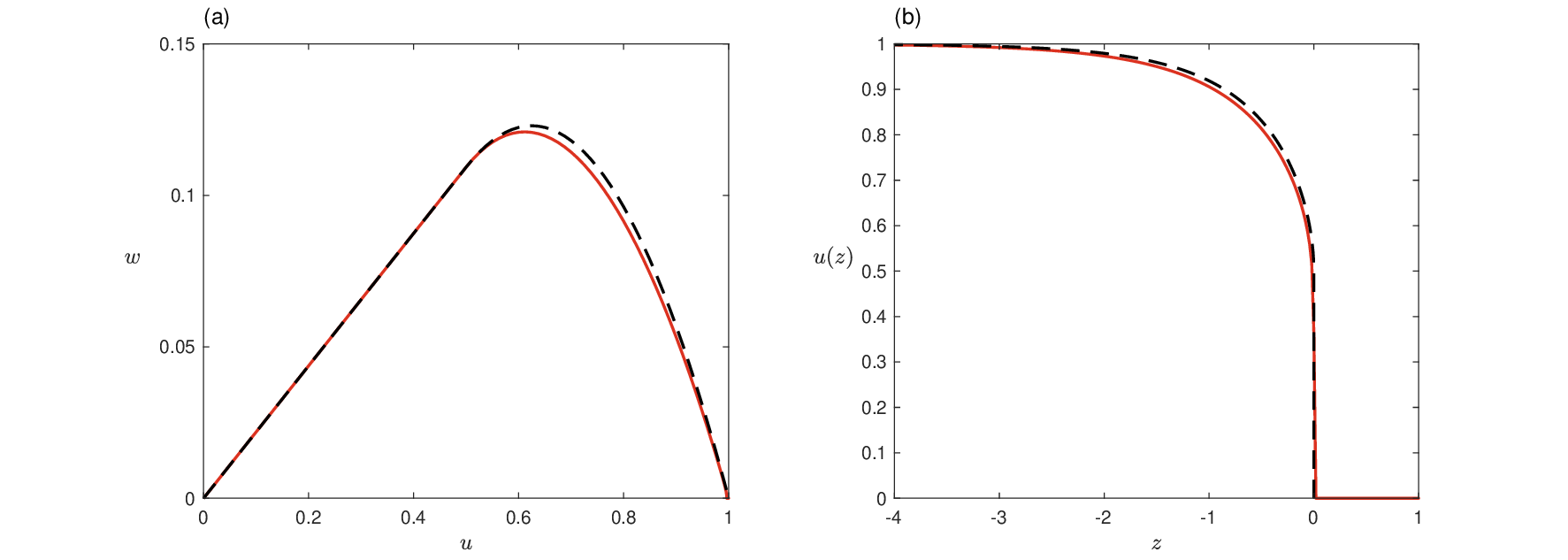}
    \caption{Comparison between numerical solution (solid line) and approximations (dashed line) for $w=Dv$ (a) and front shape (b), in the case $\theta=1/2$.}
    \label{fig:5}
\end{figure}

We can however, still obtain an approximation to the wavespeed, but not by maximising over a chosen set of trial functions. Instead we approximate the solution to the ODE \eqref{eq.2a_equiv}  by introducing $w =D v$ which satisfies 
\begin{equation}
w(w' - c) + f D = 0. 
\end{equation}
Next we write down an approximation to the trajectory  $w(u)$, and use this to generate a trial function $\tilde s(u)$ which we take as an estimate of $\hat s(u)$ in the variational scheme. In the range $u \in [0,\theta]$, $D(u)=0$ so we simply have $w = cu$ . In the range $u\in[\theta,1]$ we approximate  $w \approx c(1-u)(u-\theta^2)/(1-\theta)^2$; this is the quadratic that yields continuity of $w$ and its slope at $u=\theta$ and satisfies the condition  $w(1)=0$ forced by $v(1)=0$  (see Fig.\ref{fig:5}a). Substitution of $w$ into \eqref{eq.s_hat} gives
\begin{equation}
    \tilde{s}(u)=\exp \int \frac{c}{w} du \approx 
    \begin{cases}
    u\,\theta^{\frac{1-\theta}{1+\theta}-1}& \text{if } u \leq \theta
    \\
    \displaystyle \left(\frac{u-\theta^2}{1-u}\right)^{\frac{1-\theta}{1+\theta}} &\text{otherwise.}
    \end{cases}
\end{equation}
The resulting expression for the lower bound on wave speed is complicated. With some help from Mathematica, it may be written:
$$\frac{1}{2}c^2\geq \frac{(\theta +1)
   \left(\theta ^2-1\right)^{\frac{\theta -1}{\theta
   +1}} \left(\theta ^3 \, H_2(\theta)-\left(2 \theta ^2+\theta -1\right) \theta 
   \, H_1(\theta)+(\theta -1) (\theta
   +1)^2 \, H_3(\theta)\right)}{(\theta -1)^{-\frac{1-\theta}{1+\theta}-3}\theta
   ^{2-\frac{1-\theta}{1+\theta}} (\theta +3)-(\theta -1)^{-3} (\theta +1) \left(\theta
   ^2-1\right)^{1-\frac{1- \theta }{1+\theta}} \,
   H_1(\theta)}
$$
where 
\begin{equation}
    \begin{split}
    &H_1(\theta)=\, _2F_1\left(-\frac{2 \theta }{\theta +1},\frac{\theta
   +3}{\theta +1};2+\frac{2}{\theta +1};\frac{1}{\theta
   +1}\right)\\
    &H_2(\theta)=\, _2F_1\left(\frac{\theta +3}{\theta +1},\frac{2}{\theta
   +1}-1;2+\frac{2}{\theta +1};\frac{1}{\theta
   +1}\right)\\ 
   &H_3(\theta)=\, _2F_1\left(\frac{\theta +3}{\theta +1},\frac{2}{\theta
   +1}-3;2+\frac{2}{\theta +1};\frac{1}{\theta
   +1}\right)\,,\\
    \end{split}
\end{equation}
and $_2F_1$ is the hypergeometric function. The bound is, though, very well approximated (see Fig.\ref{fig:4}c) by the simple expression 
\begin{equation}
    c\approx\frac{1}{\sqrt{2}}(1+\theta^2)(1-\theta)^2\,.
    \label{eq.csimple}
\end{equation}

Perhaps not surprisingly, the travelling waves are sharp-fronted for all values of $\theta$.  Using our approximations for $w$ and $c$ we find a close approximation to the inverse of the wavefront, centred at zero (see Fig.\ref{fig:5}b):
$$
z=-\int \frac{D}{w}\text{d}u \approx \begin{cases}
\frac{\sqrt{2}}{(1+\theta) \left(1+\theta ^2\right)} \left(\theta  \log \left(\frac{u-\theta ^2}{\theta-\theta^2 }\right)+\log \left(\frac{1-u}{1-\theta}\right)\right)\quad&u>\theta\\
   0&\text{otherwise.}
\end{cases}
$$
As in the other cases considered, the width of the front decreases with wavespeed.

\section{Summary}

We have explored travelling wave solutions of the density-dependent FKPP equation \eqref{eq.general_DDD} for a broader range of density functions $D(u)$ than has previously been considered. Travelling waves propagate at the minimum realisable speed $c$. For most choices of $D(u)$,  $c=c_L$, the (minimum) linear speed determined by the value of $D(u)$ and the derivative of the growth term $f(u)$ as $u\to 0$. The shape of the (pulled) front is affected very little by the details of $D(u)$ for $u>0$.  Most of the key properties of solutions to \eqref{eq.general_DDD} can therefore be found through analysis of linear diffusion functions. The key condition needed for $c_L$ to be the selected speed is that $D(0)$ is sufficiently greater than zero, but finite. A consequence of the dependence of $c$ on $D(0)$ is that any behavioural mechanism that increases diffusion at low density will in principle have a strong effect on the speed of population dispersal.

If $D(0)$ is small or zero, variational methods allow us to compute minimum realisable speeds, enabling us to extend known results for $D(u)=u$ to diffusion functions with positive density dependence over a baseline $\delta$ (Section \ref{baseline}) and above a threshold $\theta$ (Section \ref{threshold}).  For the former, $c>c_L$
up to a critical value of the baseline $\delta$; wave profiles are smooth for $\delta>0$. All wave profiles in the threshold case are sharp-fronted, and the selected speeds decrease monotonically with $\theta$.

\end{document}